\documentclass{optica-article}

\journal{opticajournal} 

\articletype{Research Article}

\usepackage{lineno}

\usepackage{graphicx}
\usepackage[percent]{overpic}

\usepackage{braket}

\newcommand\eq[1]{Eq.~(\ref{eq:#1})}

\newcommand\fig[1]{Fig.\ref{fig:#1}}

\newcommand\secn[1]{Section~\ref{secn:#1}}

\begin{document}

\title{Machine learning for efficient generation of universal photonic quantum computing resources}

\author{Amanuel Anteneh,\authormark{*} L\'eandre Brunel,\authormark{*} and Olivier Pfister\authormark{$\dag$}} 
\address{University of Virginia Physics Department, 382 McCormick Rd, Charlottesville, VA 22903, USA\\
\authormark{*}These authors contributed equally to this work.}
\email{\authormark{$\dag$}olivier.pfister@gmail.com}

\begin{abstract*}
We present numerical results from simulations using deep reinforcement learning to control a measurement-based quantum processor---a time-multiplexed optical circuit sampled by photon-number-resolving detection---and find it generates squeezed cat states quasi-deterministically, with an average success rate of 98\%, far outperforming all other proposals. Since squeezed cat states are deterministic precursors to the Gottesman-Kitaev-Preskill (GKP) bosonic error code, this is a key result for enabling fault tolerant photonic quantum computing. Informed by these simulations, we also discovered a one-step quantum circuit of constant parameters that can generate GKP states with high probability, though not deterministically. 
\end{abstract*}

\section{Introduction}
Continuous-variable (CV) quantum information, as based on bosonic fields, a.k.a.\ qumodes, provides an eminently and naturally scalable platform for quantum computing (QC), as demonstrated by the generation of quantum optical Gaussian(-Wigner-function) cluster states of record sizes~\cite{
Chen2014,
Yoshikawa2016,Asavanant2019,Larsen2019} for measurement-based quantum computing~\cite{Raussendorf2001}. 
Moreover, bosonic codes such as the non-Gaussian Gottesman-Kitaev-Preskill (GKP) states~\cite{Gottesman2001} encode a qubit or qudit in an oscillator and, as such, allow fault-tolerant quantum computing over continuous variables~\cite{Menicucci2014ft}. GKP states are of fundamental importance as they are non-Gaussian resources which are required for exponential speedup~\cite{Bartlett2002}  in CVQC. 
Moreover, GKP states specifically allow the use of the Gaussian CVQC toolbox~\cite{Pfister2019} for qubit-based QC and are in fact the only non-Gaussian resource needed for universal fault tolerant CVQC~\cite{Baragiola2019}. GKP states have been experimentally realized in the phononic field that describe the vibration of a trapped-ion qubit~\cite{Fluhmann2019} and in the microwave cavity mode coupled to a superconducting qubit~\cite{CampagneIbarcq2020}. Because of the aforementioned record scalability of optical systems, the optical generation of GKP states is a crucial endeavor and has recently seen preliminary results~\cite{Konno2024}.
Different methods exist for generating GKP states, one being to couple the bosonic qumode with a qubit to perform controlled unitaries~\cite{Fluhmann2019,CampagneIbarcq2020}. In the absence of qubits, another method uses the interference of two squeezed cat states~\cite{Vasconcelos2010,Weigand2018}. Here, we show that squeezed cat states can be generated  with a 98\% average success rate by a measurement-based optical circuit driven by reinforcement learning.


\section{Machine learning for quantum optics} 
Reinforcement learning, particularly deep reinforcement learning (DRL) which utilizes deep neural networks, has proven to be a powerful approach for solving both deterministic and stochastic control tasks \cite{mnih2015human, borah2021measurement}. It consists of an agent, modeled by neural networks, interacting with an environment (the optical circuit) by selecting actions on the environment that produce a reward. Given the density matrix of the output state of the circuit as input, the goal of the agent is to learn which circuit parameters to select such that the cumulative reward (fidelity of the generated state to a squeezed cat state) is maximized. Much of the prior work on using machine learning methods for circuit optimization has primarily focused on circuits that do not include  measurements but instead utilize non-Gaussian unitaries such as Kerr interactions  \cite{arrazola2019machine, kudra2022robust}. In this case gradient-descent based approaches can be used to optimize the individual parameters of the gates in the circuit directly. In the case of \cite{tzitrin2020progress} PNR measurements were used as the non-Gaussian resource and a global optimization algorithm was used to optimize the individual gate parameters of the GBS-like device for a subset of all possible detection patterns. The disadvantage with this approach is three-pronged. Firstly, even for a 10 mode device and a maximum of 30 photons detected on any mode we have $10^{30}$ possible detection patterns to optimize over and for each of those detection patterns the chosen optimization routine must be run for some number of iterations $N$ to find the optimal parameters resulting in potentially $N\times10^{30}$ evaluations of the objective function that we seek to maximize. Since evaluation of the objective function requires execution of the classically simulated circuit this can lead to prohibitively high computation times. Secondly, once the optimal gate parameters are found for a particular detection pattern any deviation from this pattern is likely to significantly degrade the quality of the output state due to the discrete nature of the measurement basis. 
Lastly the traditional forms of the gradient-based optimization methods used in the aforementioned works must perform function evaluations sequentially as each gradient update depends on the previous one. In contrast our method allows us to take advantage of parallelization to compute many function evaluations at once. 

By recasting the measurement-based circuit optimization problem as a sequential and stochastic control task we allow for the use of DRL techniques which can learn state preparation strategies that are adaptive without the need to observe every possible detection sequence. 
\subsection{Quantum optical circuit}

The quantum optical circuit is depicted in \fig c. 
\begin{figure}[ht]
\vglue -.05in
\begin{center}
\includegraphics[width=.5\columnwidth]{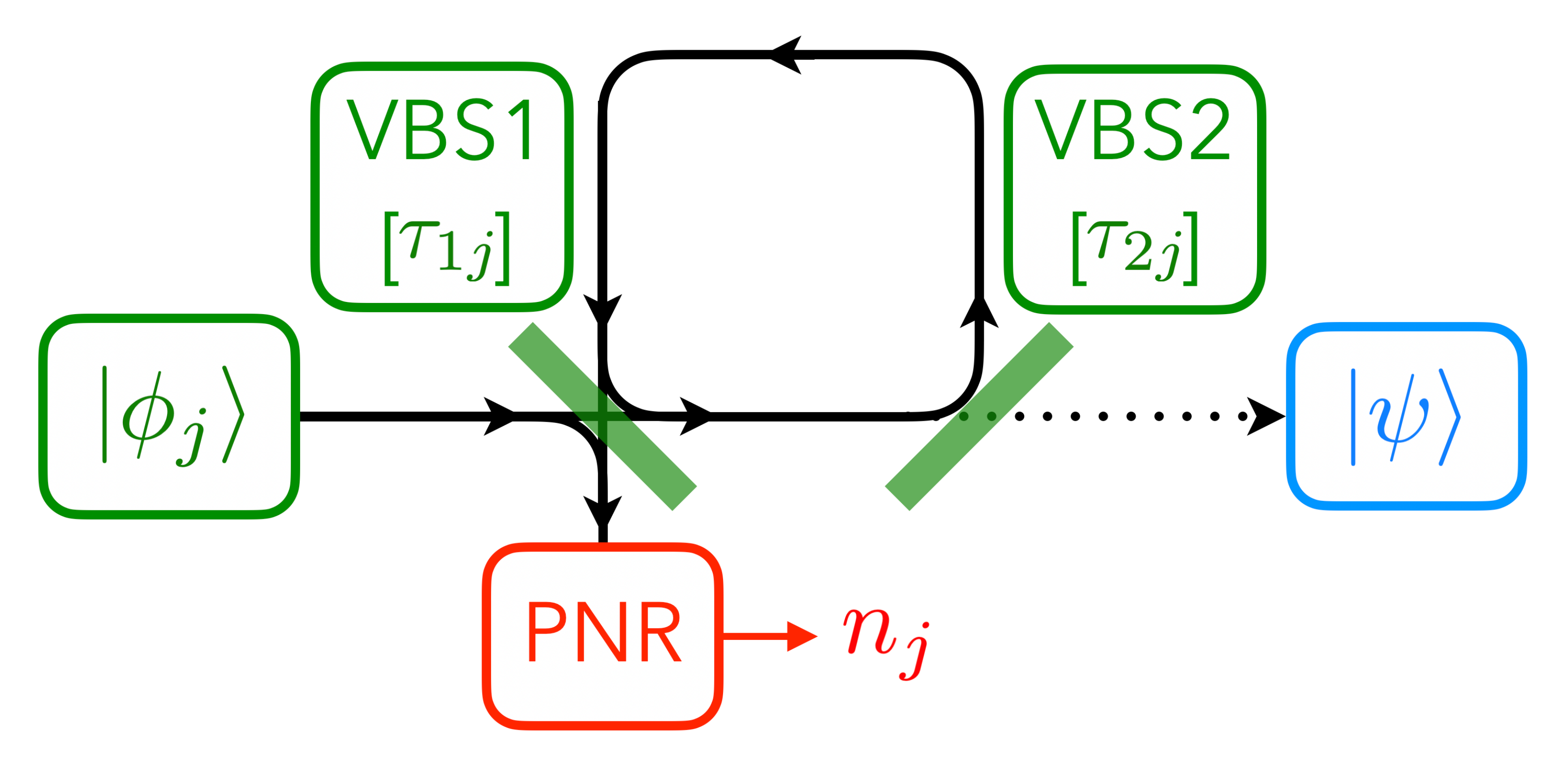}   
\end{center}
\vglue -.15in
\caption{Quantum optical circuit for squeezed cat state generation driven by deep reinforcement learning. The agent's input is shown in red and its parameters in green. The end result is in blue. VBS: variable beam-splitter; PNR: photon-number-resolving (measurement).}
\label{fig:c}
\vglue -.1in
\end{figure}
At step $j$ ($j=0,...,m$), input state $|\phi_j\rangle$ is split by VBS1, of field transmission coefficient $\tau_{1j}$
, and sent back to VBS1 via a phase-stable loop. VBS2 is a switchable mirror of $\tau_{2j}$=0 for $j$$<$$m$ and $\tau_{2m}$=1. The other output of VBS1 undergoes a photon-number-resolving (PNR) measurement yielding result $n_j$, while the loop's input to VBS1 will interfere with state $|\phi_{j+1}\rangle$. All input states are squeezed vacuum states $|0,r_j,\theta_j\rangle$ of adjustable squeezing parameter $r_j$. Note that the phase inside the loop and the squeezing angle $\theta_j$ are both set to 0 for these simulations but can be controlled by the agent as well if desired.

\subsection{Machine learning simulation}\label{secn:ml}

Reinforcement learning algorithms are designed to control Markov decision processes \cite{sutton2018reinforcement}. A Markov decision process consists of a set of states $\mathcal{S}$ which describe the state of the environment, a set of actions $\mathcal{A}$ the agent can select from to interact with the environment, a reward function $R(s_j)$ which returns a scalar reward when the state $s_j$ is arrived at and a transition function $\mathcal{T}(s_j,a_j)$ which is probabilistic in general and determines the state the environment transitions to based on the current state and the action selected by the agent. A Markov decision process must also possess the Markov property which means that the next state $s_{j+1}$ is only dependent on the current state $s_j$. The state of the environment $s_j$ at step $j$ is represented using the the density matrix of the input state $\rho_j=\ket{\psi_j}\bra{\psi_j}$. In particular we define a vector
$s_j=[\textrm{Re}(\rho_j^{u}), \textrm{Im}(\rho_j^{u}), 
\textrm{diag}(\rho_j)]$ where $\rho_j^{u}$ denotes all entries of $\rho_j$ that are above its diagonal. An action is represented as a vector $a_j=[r_j,\tau_{2j}]$. 
The reward function $R(s_j)$ is given by the fidelity of the current state $\rho_j$ with one of 4 target states
\begin{align}\label{eq:f}
    R(s_j)=\left(\max\{\mathcal{F}(\rho_j,\rho_{\pm}),\mathcal{F}[\rho_j,\mathcal{R}(\tfrac{\pi}{2})\rho_{\pm}\mathcal{R}^\dag(\tfrac{\pi}{2})]\}\right)^P 
\end{align}
where $\mathcal{R}(\tfrac{\pi}{2})$ is the $\pi/2$ rotation in quantum phase space, i.e.\ the quantum Fourier transform (FT) and the fidelity is given by~\cite{Jozsa1994}
\begin{align}
    \mathcal F(\rho_1,\rho_2)={\rm Tr^2}\left[\sqrt{\sqrt{\rho_1}\rho_2\,\sqrt{\rho_1}}\right]
\end{align}
The target states $\rho_\pm$=$\ket{\chi_\pm}\bra{\chi_\pm}$ are squeezed cat states of $\pm$ parity 
\begin{align}
\ket{\chi_\pm} \propto e^{\frac r2(a^{\dag\,2}-a^2)}\,(\ket\alpha\pm\ket{-\alpha}),
\end{align} 
$a$ being the photon annihilation operator and $a\ket\alpha$=$\alpha\ket\alpha$. The Wigner functions of these 4 states are plotted in \fig t. Note that $\alpha>1$ is required, $\forall r$, so that the lobes are clearly separated from the fringes. 
\begin{figure}
    \centering
    \includegraphics[width=.6\columnwidth]{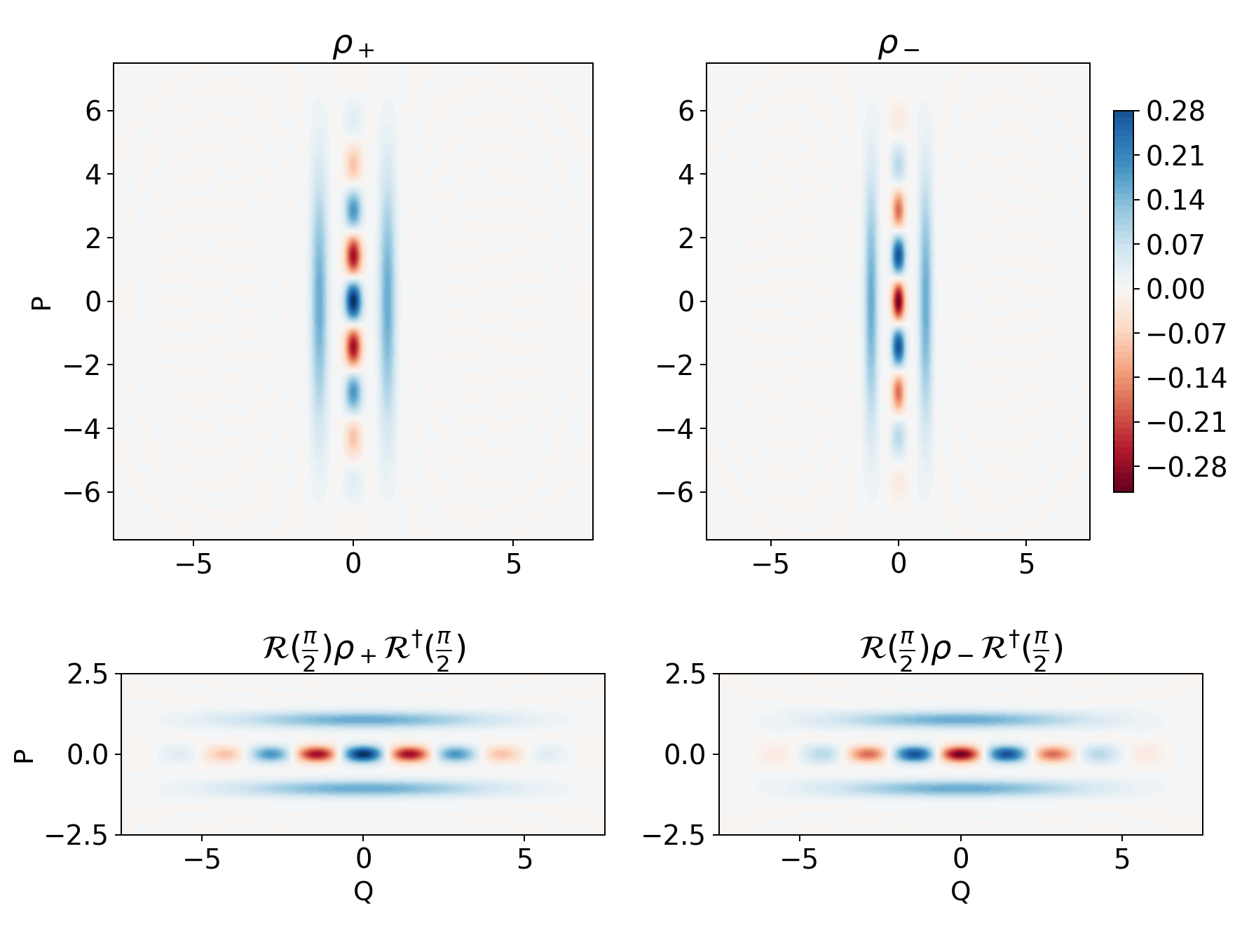}
    \caption{Wigner functions of the 4 target squeezed cat states ($r$=1.38, $\alpha$=3) used in the reward function calculation of \eq f.}
    \label{fig:t}
\end{figure}
Each of the 4 target states can be distinguished by the detection sequence of their generation episode: even-parity states (positive Wigner function at the origin) are generated when the total detected photon number $N$=$\sum_{j=0}^m n_j$ is even. Odd-parity states (negative Wigner function at the origin) have odd $N$. Moreover, FT'd states can be distinguished by checking the agent's action in the last step of the episode: FT'd states have $r_m$$<$0, the other have $r_m$$>$0. Exponent $P$ in \eq f is used to penalize low-fidelity states~\cite{porotti2022deep} and is taken to be 50 in our simulations. 
\begin{figure*}[ht]
\vglue -.05in
\begin{center}
\includegraphics[width=.8\linewidth]{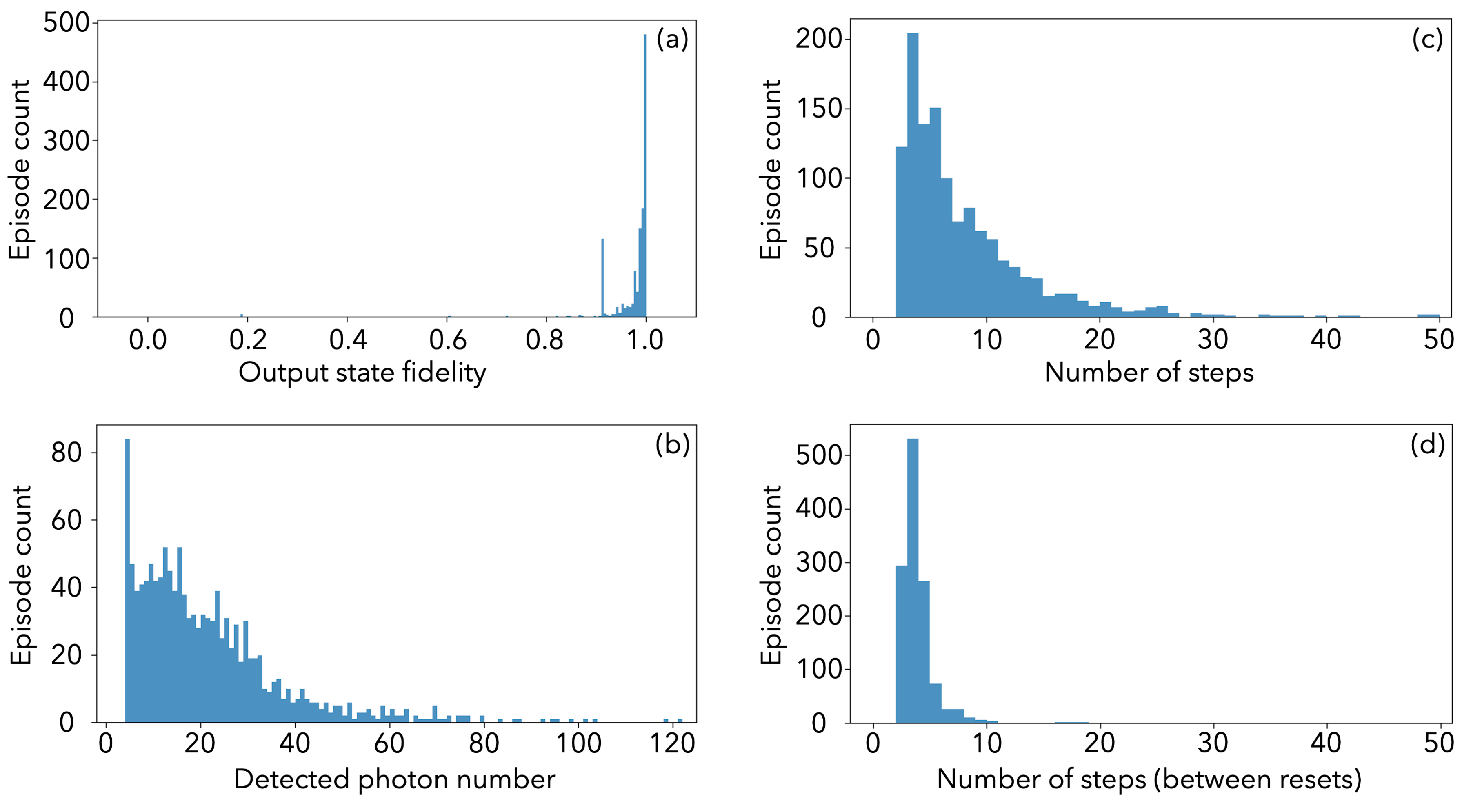}   
\end{center}
\vglue -.15in
\caption{Histograms for 1250 generation episodes ($m=50$).(a), output state fidelity; (b), detect photon number per episode; (c), steps per episode, with resets; (d), steps per episode, between resets.}
\label{fig:r}
\vglue -.1in
\end{figure*}
The initial state $\ket{0,r_0}$ has $r_0$=1.38. We used Python libraries \texttt{StableBaselines3}~\cite{raffin2021stable} to implement proximal policy optimization (PPO)
as our reinforcement learning algorithm and \texttt{StrawberryFields}~\cite{killoran2019strawberry} to simulate the quantum optical circuit. We utilize an actor-critic version of PPO so our agent consists of two deep neural networks, an actor network which selects an action given a particular state $s_j$ and a critic network which approximates the value $V(s_j)$ of a state \cite{sutton2018reinforcement}.  Each network consists of three hidden layers of size 256, 128 and 64 respectively with $\tanh$ as the activation function. We used 40 separate circuit environments with different random seeds to speed up collection of samples to train the neural network on.

\subsection{Results}

\subsubsection{Lossless case}
The agent was trained over 7$\times$$10^6$ 
steps with 10-step episodes. We evaluated the results by examining 5 sets of 250 50-step episodes, each set having a different random-number seed to simulate PNR results. Each episode is composed of at most 50 steps, i.e., loop iterations, and the agent invariably concluded each and every episode by setting $\tau_{1j}$, denoted as $\tau_{j}$ from here on, to 0 and not changing it in all steps thereafter. \fig r captures the results. \fig r(a) clearly demonstrates high fidelity and further scrutiny of each individual episode revealed that the outcome is a squeezed cat state 98\% of the time on average, be it different from the target state at times. In assessing success, we therefore included all obtained squeezed cat states, even those with parameters different from ($r$=1.38,$\alpha$=3) and therefore suffering from lower fidelities, for the following two reasons: {\it(i)}, it is well known that small cat states can be deterministically ``bred'' to larger sizes~\cite{Lund2004,Etesse2015,Sychev2017} so they are still useful and, {\it(ii)}, all generation sequences $\{$($r_j$,$\tau_j$,$n_j$)$\}$$_j$ uniquely label each output which allows generation of accurate lookup tables to drive an experimental setup with full knowledge of the generated state based on the generation sequence. These considerations validate the observed average success rate of 98\% for 50-step episodes. 

\fig r(b) gives a sense of the average detected photon number per episode, which includes large detected photon numbers at times as the number of steps per episode was often much less than 50, as \fig r(c) shows. 

\subsubsection{Lossy case}
The simulation was also run with a 1\% loss channel placed just before the PNR detector, yielding a 75\% success rate for 4.5$\times$$10^6$ training steps with no degradation of the state quality. When the loss was increased to 10\%, the agent was unable to find success and always stopped at step one. We believe more training should improve on the 1\% results and conjecture that considerably more training might also allow the 10\% case to become successful. How much more training would be needed is hard to quantify. The current training levels are all at the limit of our current computing capabilities.

\begin{figure}[ht]
\vglue -.05in
\begin{center}
\includegraphics[width=.7\columnwidth]{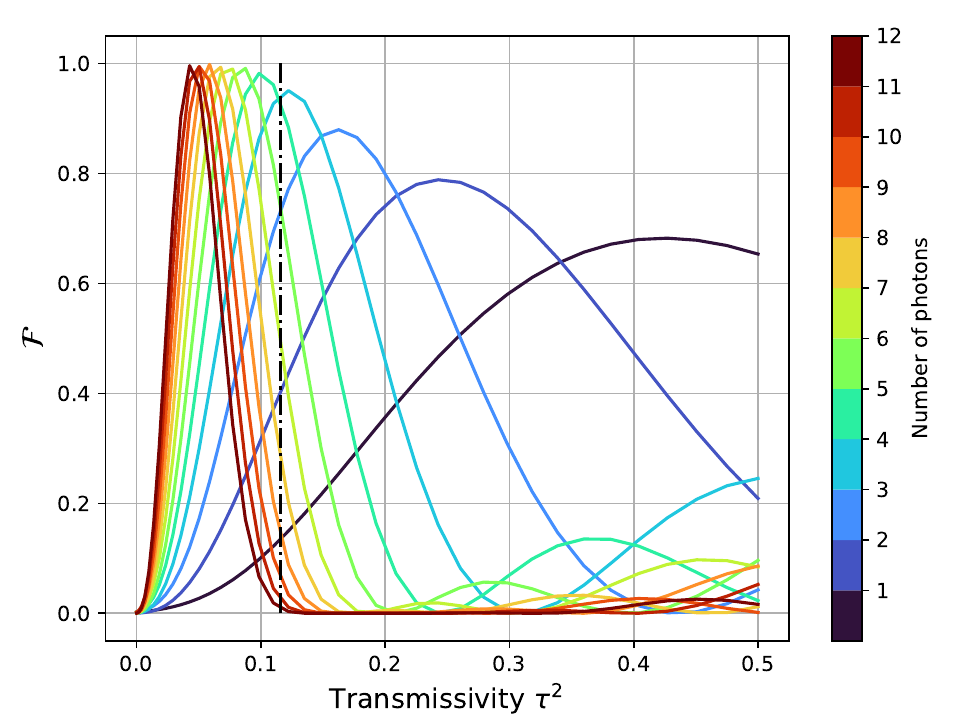}   
\end{center}
\vglue -.15in
\caption{Fidelities of the output state of the first step with a target squeezed cat state with $(\alpha,r)=(3,1.38)$. The vertical dot-dashed line marks the initial transmittivity $\tau_1^2$=0.12 ($\tau_1$=0.34).}
\label{fig:alpha_r_fixed}
\vglue -.1in
\end{figure}

\subsubsection{State generation dynamics}

Turning now back to the lossless case, we made two important observations: first, the agent often ``reset'' the episode by setting $\tau_{j}$=1, i.e., flushing out the optical loop and starting afresh with a squeezed state. Second, we then noticed that very few steps were required between resets, \fig r(d), in contrast to the efficient PhANTM cat generation method~\cite{Eaton2022_PhANTM}. This observation prompted us to investigate single-step state generation based on the strategy learned by the agent. While this single-step protocol cannot rival with the ML episodes, it is still new and interesting to develop. 

\section{Machine-learning-informed, fixed-parameter, single-step strategies}

Two conclusions can be immediately drawn from the results of the previous section: \textit{(i)}, when ``lucky,'' the trained agent only needs $3\pm1$ steps to be successful, \fig r(d); \textit{(ii)}, the trained agent always chooses specific steps to carry out its task, case in point by always setting $\tau_1$ = 0.34. The latter observation, {\it(ii)}, is to be expected since the agent will always chose the same action given the same state since a deterministic version of the policy learned during training is used during inference. Finding {\it(i)} was surprising, however, as our initial expectation of ML was that it would enable us to navigate a vast parameter space over a number of steps much larger than 3 by sorting through large numbers of random photon number sequences $\{n_j\}_{j\in[1,m]}$. For this particular task of generating squeezed cat states, it wasn't the case.

One might then be inclined to think that such a few-step evolution could be easily modeled with analytic tools such as Kraus operators. However, an analytic theoretical study of the quantum circuit evolution for arbitrary parameters in \fig c turned out to be actually quite complicated even in the single-step case, enough to justify relying on numerical simulations to tackle this problem. 
In \secn{fixed}, we investigate the single-step behavior of the circuit for a variable $\tau_1=\tau$, a detected photon number $n_1=n$, and a fixed-target squeezed cat state $(\alpha,r)=(3, 1.38)$. In \secn{moving}, we use a moving-target squeezed cat state.

\subsection{One-step generation of a fixed-target squeezed cat state $(\alpha,r)=(3, 1.38)$} \label{secn:fixed}

Considering the first step of our protocol and by changing the transmittivity of the beam splitter, we first analyse the variation of the fidelity of the resulting state with the fixed target squeezed cat state $(\alpha,r)=(3, 1.38)$ presented in  \fig t.
\fig{alpha_r_fixed} displays the fidelities $\mathcal{F}_{n}$ of the one-step output state with the fixed target state, depending on $n$ and $\tau$. We discard the case $n=0$, which cannot yield a non-Gaussian output state---and will still take these outcomes into account when evaluating success probability. The good news is that very high fidelities are obtained for a wide range of values of $n$; the bad news is that these high values are obtained for different optimal values of $\tau$, unsurprisingly. Since $n$ is a random variable, the choice of $\tau$ cannot be informed directly by \fig{alpha_r_fixed}.

\begin{figure}[ht]
    \centering
    \includegraphics[scale=0.4]{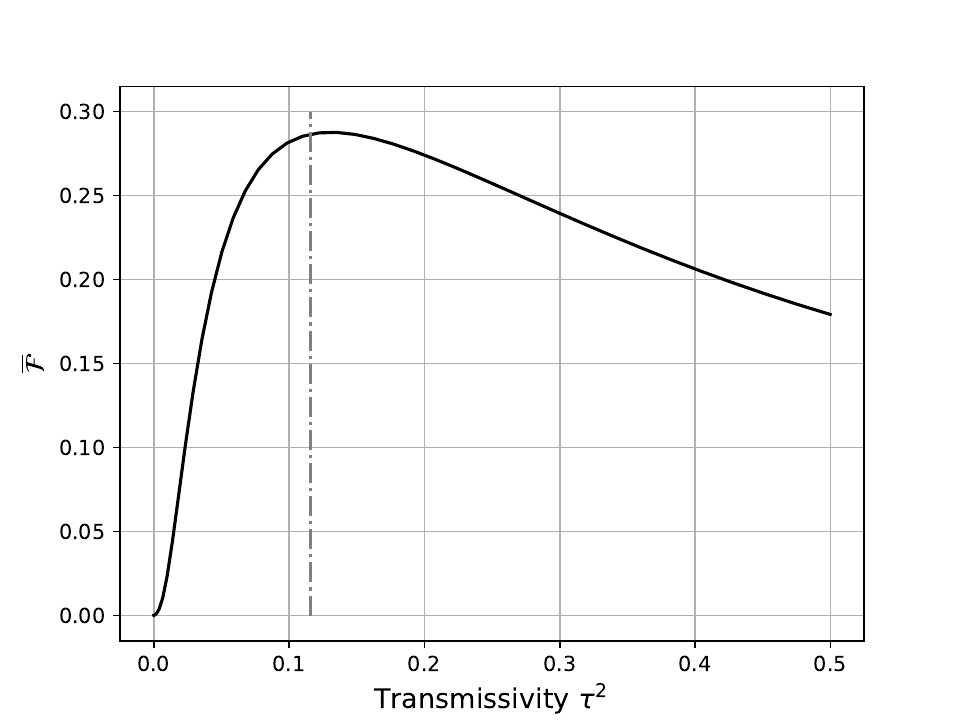}
    \caption{Average fidelity versus transmissivity. The dot-dashed line marks the initial transmissivity ($\tau=0.34$) chosen by the trained agent.}
    \label{fig:mean_fid}
\end{figure}

In order to refine this analysis, we elected to compute the average fidelity over probability distribution $p_{n}$ of $n$,
\begin{align}
    \overline{\mathcal{F}}(\tau) = \sum^{\infty}_{n=1} p_{n}\mathcal{F}_{n}(\tau),
\end{align}
plotted in \fig{mean_fid}. As can be seen, the initial transmissivity chosen by the trained agent is very close to the optimal average fidelity found at $\tau^2 = 0.135$ ($\tau = 0.367$). 

Turning now to the success probability of a single-step circuit, one could certainly seek an optimum for $\tau$ using other criteria than the average fidelity, but a natural question emerges at this point: what if the single step produced good-quality, hence desirable, squeezed cat states with  $(\alpha,r)\neq(3,1.38)$ and possibly poor overlap with the target state \footnote{Intuitively, the cat fringes are non-Gaussian features of the Wigner function and have period $\xi=\frac{\pi e^{2r}}{4 \alpha}$. These fringes will overlap well for squeezed cat states whose parameters $(\alpha,r)$ give the same ratio $\xi$.}? 
In order to assess this case, we decided to ``move the goal posts'' and study the single-step circuit with a moving-target state.


\subsection{One-step generation of a moving-target squeezed cat state}\label{secn:moving}

\begin{figure}[ht]
\vglue -.05in
\begin{center}
\includegraphics[width=.7\linewidth]{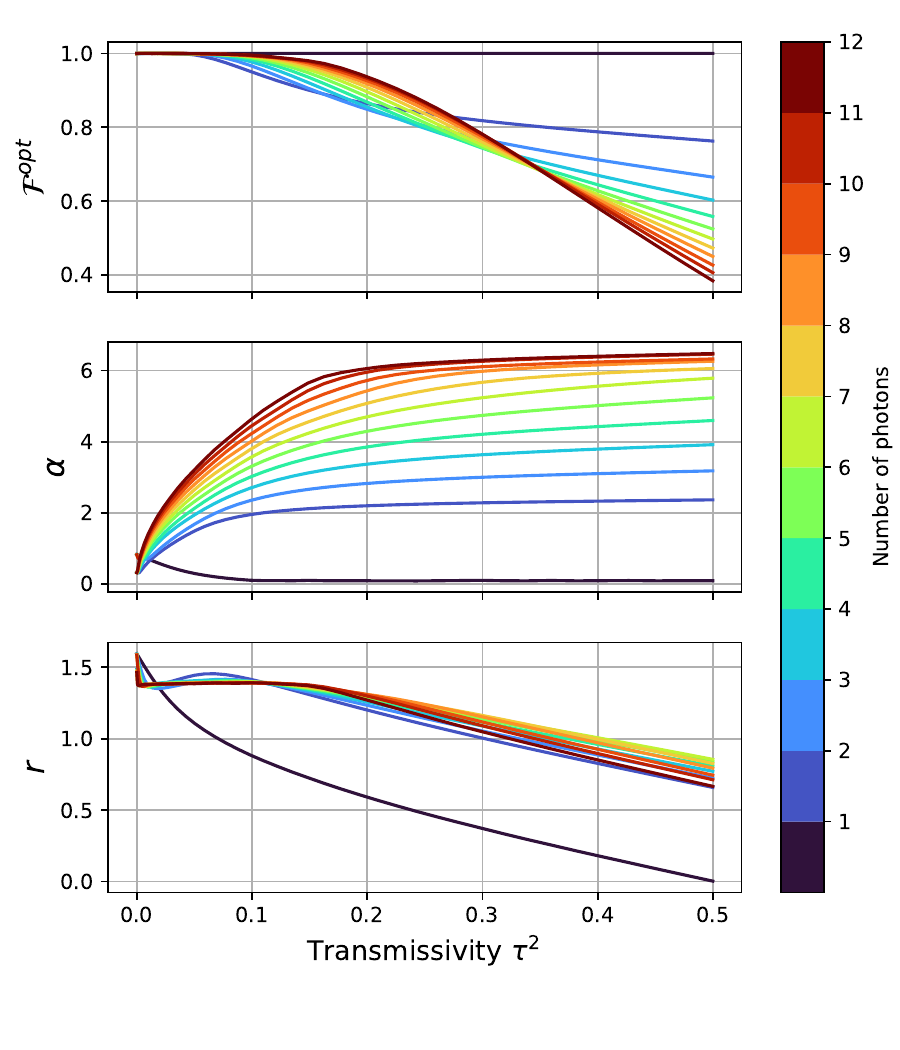} 
\end{center}
\vglue -.4in
\caption{Top, optimal fidelity with a target state of adjustable $\alpha$ and $r$. Middle, corresponding values of $\alpha$. Bottom, corresponding values of $r$.}
\label{fig:all_param_opt}
\end{figure}

For each value of $\tau$ and $n$, we optimized the fidelity of the output state with a squeezed cat state of variable parameters $\alpha$ and $r$. The results for $\mathcal F_n^\text{opt}(\tau)$, $\alpha_n(\tau)$, and $r_n(\tau)$ are displayed in \fig{all_param_opt}.
Comparing with \fig{alpha_r_fixed}, we immediately see that $\mathcal F_n^\text{opt}(\tau)$ is unity for a wider range of $\tau^2$. However, the lowest values of $\tau^2$ give states which have low $\alpha\simeq1$ and are  not well-resolved cats, and therefore not desirable. In fact, $n=1$ never gives $\alpha>1$ even though $\mathcal F_1^\text{opt}(\tau)=1$, $\forall \tau$. For $n>1$, there is a ``sweet spot'' for $\tau^2$, between 0.1 and 0.2, for which $\mathcal F_n^\text{opt}(\tau)$ is near unity and $\alpha$ and $r$ are large. We can also see that the optimal values of $\tau$ found in \fig{alpha_r_fixed} are confirmed in \fig{all_param_opt}. 

\begin{figure}[ht]
\centering
\includegraphics[scale=0.4]{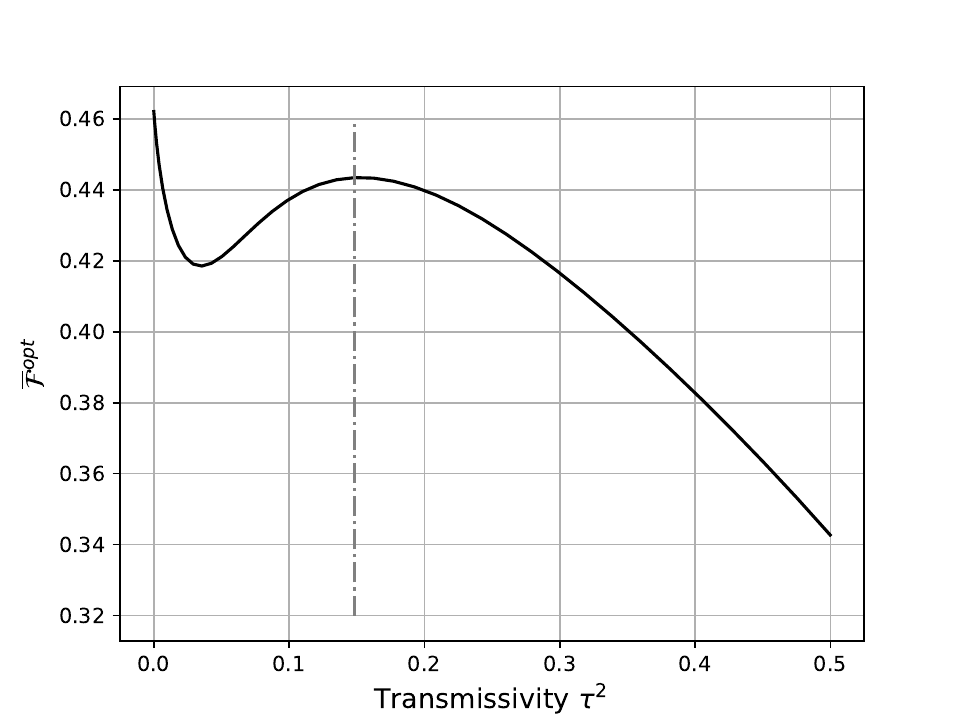}
    \caption{Average optimal fidelity versus transmissivity. The dot-dashed line marks the relevant secondary maximum at $\tau^2=0.146$.} 
    \label{fig:mean_fid_move_r_alpha}
\end{figure}

Here again, we calculated the average fidelity 
\begin{align}
    \overline{\mathcal F}^\text{opt}(\tau) = \sum^{\infty}_{n=1} p_{n}\mathcal{F}_{n}^\text{opt}(\tau),
\end{align}
which is plotted in \fig{mean_fid_move_r_alpha}.
We see that $\overline{\mathcal F}^\text{opt}(\tau)$ reaches larger values than $\overline{\mathcal F}(\tau)$ before (\fig{mean_fid}). As mentioned previously, we discard the region near $\tau=0$ which correspond to low-quality states. The relevant optimum is then found at $\tau^2=0.146$ ($\tau=0.382$), again near to the agent's chosen value of 0.12. We will adopt this value $\tau^2=0.146$ in the GKP circuit described in the next section.

This study confirms that large fidelities are reachable for squeezed cat states outside our initial choice of $(\alpha,r)=(3, 1.38)$ for the target state. This was observed in some output states generated by the agent, which were high-quality squeezed cat states but had poor overlap with the arbitrarily chosen target. This could, in principle, be remedied by refining the reward function, though likely at the cost of longer training times.

At this point, a simple single-step circuit of \fig{c} followed by cat/GKP state breeding protocols can be used to created GKP states with reasonable success probability. 

\subsection{One-step generation of GKP states} 

\begin{figure*}[ht]
\vglue -.05in
\centerline{
\includegraphics[width=\linewidth]{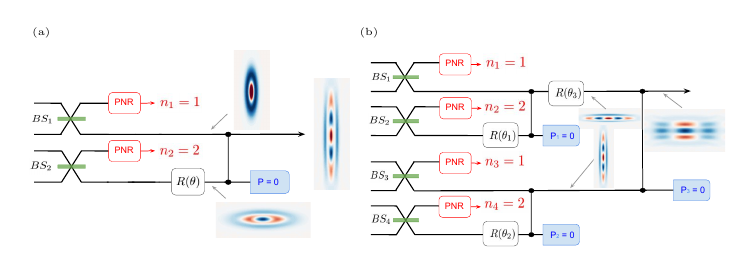}   
}
\vglue -.3in
\caption{Cat~\cite{Eaton2022_PhANTM} and GKP breeding protocols. All input states consist in orthogonal squeezed quadratures into beamsplitters (BS) of transmissivity $\tau^2 = 0.146$. Depending on the measured photon number $n$ on the PNR detectors, the output squeezed cats can be bred to larger cats, (a), and then to GKP states, (b). The vertical black line between horizontal optical channels denotes a continuous-variable CZ gate. Depending on the PNR measurement outcomes, the user can choose to teleport through a $C_Z$ gate to get a bigger cat, by rotating the bottom state by $\theta = \frac{\pi}{2}$. (a), Cat breeding protocol for $(n_1,n_2)=(1,2)$. The prepared input state results from the PNR measurement $n = 1$ photon, which overlaps with fidelity $\mathcal{F}_1$ = 0.99  with a squeezed cat state with $\alpha^{opt}_1 = 0.10$ and $r^{opt}_1 = 0.73$, and from the PNR measurement $n_2 = 2$ photons, which overlaps with fidelity $\mathcal{F}_2$ = 0.90 with a squeezed cat state with $\alpha^{opt}_2 = 2.12$ and $r^{opt}_2 = 1.31$. The breeding result overlaps with fidelity 0.98 with a squeezed cat state with $\alpha = 2.40$ and $r = 1.62$, which bears some resemblance to an $n=3$ preparation outcome, but with higher fidelity to a squeezed cat state. (b), GKP breeding. After performing the photon number measurement each PNRs, the user can decide to breed by teleportation through a $C_Z$ gate the outputs by setting $\theta_3$ to $\frac{\pi}{2}$. The breeding result has 0.99 fidelity with a $\frac{1}{\sqrt{2}}(\ket{\bar0} + i\ket{\bar1})$ standard GKP state with 6.25 dB squeezing, after applying 4.76 dB of squeezing (to yield the correct peak spacing in the Wigner function) and a $-\frac{\pi}{4}$ phase shift (to rotate the Wigner function appropriately in phase space). }
\label{fig:merged_protocol}
\vglue -.1in
\end{figure*}

We used the \textit{deterministic} cat and GKP breeding protocols described in Ref.~\cite{Eaton2022_PhANTM}, based on the original proposals of Refs.~\cite{Vasconcelos2010,Weigand2018} for GKP breeding. The proposed circuits are depicted in \fig{merged_protocol} and include the circuit of \fig c. 
The protocols of Ref.~\cite{Eaton2022_PhANTM} employ continuous-variable cluster states~\cite{Pfister2019} to which the output states are fused here, but they are also realizable with beamsplitters~\cite{Vasconcelos2010,Weigand2018}. In \fig{merged_protocol}(a), we show a cat breeding stage that is useful for $n\leqslant2$. If the measured photon numbers are higher, larger cat states are produced and cat breeding may not be necessary. In that case, the produced cat states may be sent directly to a GKP breeding stage. One can thus envision a PNR-informed switching of the input cat states to implement this. The fail mode of such a reconfigurable input for the cat breeding stage requires that the 4-PNR register records 3 zero-counts or more, which has a 2.2\% probability, or 2 zero-counts and at least 1 one-count, which has a 1.4\% probability. The rest of the cases do not preclude GKP state generation but still require extensive study of the details and optimization of the  algorithm, which is left for future work.

\section{Conclusion}

We presented a near-deterministic, machine-learning-driven method for generating GKP state precursors. Our all-optical method does not require coupling to a qubit and its 98\% success rate, and gate scarcity, significantly exceed that of  other qubitless  methods~\cite{arrazola2019machine,tzitrin2020progress,takase2023gottesman,Yao2023,Konno2024}, to the notable exception of the recent one by Winnel et al.~\cite{Winnel2024}, which  proposed two schemes to obtain cat states deterministically or quasi-deterministically. However, these require Fock states as well as  photon-number-resolving measurements as non-Gaussian initial resources, whereas our work only requires PNR detection.

Analyzing the machine learning outcomes, we also discovered less efficient but single-step circuits to generate GKP states.

These results open up very efficient non-Gaussian photonic resource generation, a key result for enabling fault tolerant photonic quantum computing.

\begin{backmatter}
\bmsection{Funding}
We acknowledge support from NSF grants PHY-2112867 and ECCS-2219760. 

\bmsection{Acknowledgments}
We  thank U. of Virginia Research Computing for providing access to the Rivanna computing cluster. We thank Nicholas Hayeck, Miller Eaton, Carlos González-Arciniegas, and Hussain Zaidi for their input in the preliminary stages of this work, and Jason Jabbour and Riccardo Porotti for stimulating discussions. 

\bmsection{Disclosures}
The authors declare no conflicts of interest.

\bmsection{Data Availability Statement}
Data pertaining to this paper's results may be obtained from the authors upon reasonable request. Computer code used to generate the data presented may be obtained from the authors upon reasonable request.

\end{backmatter}


\begin{thebibliography}{10}
\newcommand{\enquote}[1]{``#1''}

\bibitem{Chen2014}
M.~Chen, N.~C. Menicucci, and O.~Pfister, \enquote{Experimental realization of
  multipartite entanglement of 60 modes of a quantum optical frequency comb,}
  {\protect\JournalTitle{Phys. Rev. Lett.}} \textbf{112}, 120505 (2014).

\bibitem{Yoshikawa2016}
J.-i. Yoshikawa, S.~Yokoyama, T.~Kaji, \emph{et~al.}, \enquote{Invited article:
  Generation of one-million-mode continuous-variable cluster state by unlimited
  time-domain multiplexing,} {\protect\JournalTitle{APL Photonics}} \textbf{1},
  060801 (2016).

\bibitem{Asavanant2019}
W.~Asavanant, Y.~Shiozawa, S.~Yokoyama, \emph{et~al.}, \enquote{Generation of
  time-domain-multiplexed two-dimensional cluster state,}
  {\protect\JournalTitle{Science}} \textbf{366}, 373--376 (2019).

\bibitem{Larsen2019}
M.~V. Larsen, X.~Guo, C.~R. Breum, \emph{et~al.}, \enquote{Deterministic
  generation of a two-dimensional cluster state,}
  {\protect\JournalTitle{Science}} \textbf{366}, 369--372 (2019).

\bibitem{Raussendorf2001}
R.~Raussendorf and H.~J. Briegel, \enquote{A one-way quantum computer,}
  {\protect\JournalTitle{Phys. Rev. Lett.}} \textbf{86}, 5188 (2001).

\bibitem{Gottesman2001}
D.~Gottesman, A.~Kitaev, and J.~Preskill, \enquote{Encoding a qubit in an
  oscillator,} {\protect\JournalTitle{Phys. Rev. A}} \textbf{64}, 012310
  (2001).

\bibitem{Menicucci2014ft}
N.~C. Menicucci, \enquote{Fault-tolerant measurement-based quantum computing
  with continuous-variable cluster states,} {\protect\JournalTitle{Phys. Rev.
  Lett.}} \textbf{112}, 120504 (2014).

\bibitem{Bartlett2002}
S.~D. Bartlett, B.~C. Sanders, S.~L. Braunstein, and K.~Nemoto,
  \enquote{Efficient classical simulation of continuous variable quantum
  information processes,} {\protect\JournalTitle{Phys. Rev. Lett.}}
  \textbf{88}, 097904 (2002).

\bibitem{Pfister2019}
O.~Pfister, \enquote{Continuous-variable quantum computing in the quantum
  optical frequency comb,} {\protect\JournalTitle{Journal of Physics B: Atomic,
  Molecular and Optical Physics}} \textbf{53}, 012001 (2020).

\bibitem{Baragiola2019}
B.~Q. Baragiola, G.~Pantaleoni, R.~N. Alexander, \emph{et~al.},
  \enquote{All-{Gaussian} universality and fault tolerance with the
  {Gottesman}-{Kitaev}-{Preskill} code,} {\protect\JournalTitle{Phys. Rev.
  Lett.}} \textbf{123}, 200502 (2019).

\bibitem{Fluhmann2019}
C.~Fl{\"u}hmann, T.~L. Nguyen, M.~Marinelli, \emph{et~al.}, \enquote{Encoding a
  qubit in a trapped-ion mechanical oscillator,}
  {\protect\JournalTitle{Nature}} \textbf{566}, 513--517 (2019).

\bibitem{CampagneIbarcq2020}
P.~Campagne-Ibarcq, A.~Eickbusch, S.~Touzard, \emph{et~al.}, \enquote{Quantum
  error correction of a qubit encoded in grid states of an oscillator,}
  {\protect\JournalTitle{Nature}} \textbf{584}, 368--372 (2020).

\bibitem{Konno2024}
S.~Konno, W.~Asavanant, F.~Hanamura, \emph{et~al.}, \enquote{Logical states for
  fault-tolerant quantum computation with propagating light,}
  {\protect\JournalTitle{Science}} \textbf{383}, 289--293 (2024).

\bibitem{Vasconcelos2010}
H.~M. Vasconcelos, L.~Sanz, and S.~Glancy, \enquote{All-optical generation of
  states for ``{E}ncoding a qubit in an oscillator'',}
  {\protect\JournalTitle{Opt. Lett.}} \textbf{35}, 3261--3263 (2010).

\bibitem{Weigand2018}
D.~J. Weigand and B.~M. Terhal, \enquote{Generating grid states from
  schr\"odinger-cat states without postselection,} {\protect\JournalTitle{Phys.
  Rev. A}} \textbf{97}, 022341 (2018).

\bibitem{mnih2015human}
V.~Mnih, K.~Kavukcuoglu, D.~Silver, \emph{et~al.}, \enquote{Human-level control
  through deep reinforcement learning,} {\protect\JournalTitle{nature}}
  \textbf{518}, 529--533 (2015).

\bibitem{borah2021measurement}
S.~Borah, B.~Sarma, M.~Kewming, \emph{et~al.}, \enquote{Measurement-based
  feedback quantum control with deep reinforcement learning for a double-well
  nonlinear potential,} {\protect\JournalTitle{Physical review letters}}
  \textbf{127}, 190403 (2021).

\bibitem{arrazola2019machine}
J.~M. Arrazola, T.~R. Bromley, J.~Izaac, \emph{et~al.}, \enquote{Machine
  learning method for state preparation and gate synthesis on photonic quantum
  computers,} {\protect\JournalTitle{Quantum Science and Technology}}
  \textbf{4}, 024004 (2019).

\bibitem{kudra2022robust}
M.~Kudra, M.~Kervinen, I.~Strandberg, \emph{et~al.}, \enquote{Robust
  preparation of wigner-negative states with optimized snap-displacement
  sequences,} {\protect\JournalTitle{PRX Quantum}} \textbf{3}, 030301 (2022).

\bibitem{tzitrin2020progress}
I.~Tzitrin, J.~E. Bourassa, N.~C. Menicucci, and K.~K. Sabapathy,
  \enquote{Progress towards practical qubit computation using approximate
  gottesman-kitaev-preskill codes,} {\protect\JournalTitle{Physical Review A}}
  \textbf{101}, 032315 (2020).

\bibitem{sutton2018reinforcement}
R.~S. Sutton and A.~G. Barto, \emph{Reinforcement learning: An introduction}
  (MIT press, 2018).

\bibitem{Jozsa1994}
R.~Jozsa, \enquote{Fidelity for mixed quantum states,}
  {\protect\JournalTitle{Journal of Modern Optics}} \textbf{41}, 2315 (1994).

\bibitem{porotti2022deep}
R.~Porotti, A.~Essig, B.~Huard, and F.~Marquardt, \enquote{Deep reinforcement
  learning for quantum state preparation with weak nonlinear measurements,}
  {\protect\JournalTitle{Quantum}} \textbf{6}, 747 (2022).

\bibitem{raffin2021stable}
A.~Raffin, A.~Hill, A.~Gleave, \emph{et~al.}, \enquote{Stable-baselines3:
  Reliable reinforcement learning implementations,} {\protect\JournalTitle{The
  Journal of Machine Learning Research}} \textbf{22}, 12348--12355 (2021).

\bibitem{killoran2019strawberry}
N.~Killoran, J.~Izaac, N.~Quesada, \emph{et~al.}, \enquote{Strawberry fields: A
  software platform for photonic quantum computing,}
  {\protect\JournalTitle{Quantum}} \textbf{3}, 129 (2019).

\bibitem{Lund2004}
A.~P. Lund, H.~Jeong, T.~C. Ralph, and M.~S. Kim, \enquote{Conditional
  production of superpositions of coherent states with inefficient photon
  detection,} {\protect\JournalTitle{Phys. Rev. A}} \textbf{70} (2004).

\bibitem{Etesse2015}
J.~Etesse, M.~Bouillard, B.~Kanseri, and R.~Tualle-Brouri,
  \enquote{Experimental generation of squeezed cat states with an operation
  allowing iterative growth,} {\protect\JournalTitle{Phys. Rev. Lett.}}
  \textbf{114}, 193602 (2015).

\bibitem{Sychev2017}
D.~V. Sychev, A.~E. Ulanov, A.~A. Pushkina, \emph{et~al.}, \enquote{Enlargement
  of optical {Schr{\"o}dinger}'s cat states,} {\protect\JournalTitle{Nat.
  Photon.}} \textbf{11}, 379--382 (2017).

\bibitem{Eaton2022_PhANTM}
M.~Eaton, C.~Gonz{\'{a}}lez-Arciniegas, R.~N. Alexander, \emph{et~al.},
  \enquote{Measurement-based generation and preservation of cat and grid states
  within a continuous-variable cluster state,}
  {\protect\JournalTitle{{Quantum}}} \textbf{6}, 769 (2022).

\bibitem{takase2023gottesman}
K.~Takase, K.~Fukui, A.~Kawasaki, \emph{et~al.},
  \enquote{Gottesman-kitaev-preskill qubit synthesizer for propagating light,}
  {\protect\JournalTitle{npj Quantum Information}} \textbf{9}, 98 (2023).

\bibitem{Yao2023}
Y.~Yao, F.~Miatto, and N.~Quesada, \enquote{On the design of photonic quantum
  circuits,} {\protect\JournalTitle{arXiv:2209.06069}}  (2022).

\bibitem{Winnel2024}
M.~S. Winnel, J.~J. Guanzon, D.~Singh, and T.~C. Ralph, \enquote{Deterministic
  preparation of optical squeezed cat and {Gottesman-Kitaev-Preskill} states,}
  {\protect\JournalTitle{arXiv:2311.10510}}  (2023).

\end{thebibliography}
\end{document}